\documentstyle[12pt,epsfig]{article}
\begin{document}
\parskip=1mm

\begin{center}
{\Large \bf 
Calculating hadronic properties in strong QCD~}\footnote{ Talk given at the
Second Workshop on ELFE Physics, St Malo, France, September 1996}
\vspace*{1cm} \\
{\large \bf M.R. Pennington}
\vspace*{0.5cm} \\
Centre for Particle Theory, University of Durham, 
\vspace*{0mm}\\
Durham DH1 3LE, U.K.
\vspace*{0.2cm} \\
\end{center}
\baselineskip=5.5mm

This talk gives a brief review of the progress that has been made in 
calculating the properties of hadrons in strong QCD. In keeping with this
meeting I will concentrate on those properties that can be studied with
electromagnetic probes.
Though perturbative QCD is highly successful, it only applies in a limited 
kinematic regime, where hard scatterings occur, and the quarks move in the 
interaction region as if they are free, pointlike objects. However, the
bulk of strong interactions are governed by the long distance regime, where
the strong interaction is strong. It is this regime of length scales of the 
order of a fermi, that determines the spectrum of light hadrons and their 
properties. The calculation of these properties requires an understanding of
non-perturbative QCD, of confinement and chiral symmetry breaking.
\vspace{-4mm}
\begin{figure}[hb]
\begin{center}
\mbox{~\epsfig{file=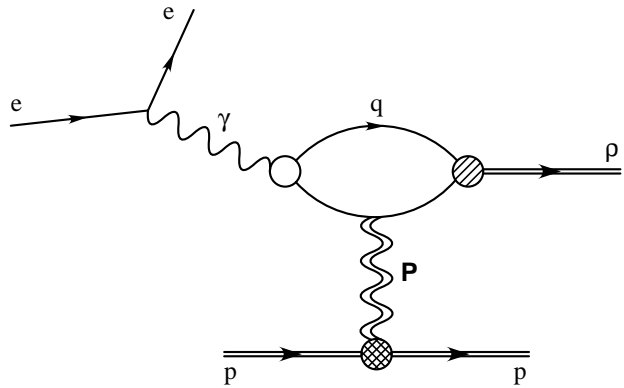,height=5.3cm}}
%\vspace*{6.5cm}
\end{center}
\vspace{-4mm} 
\caption[Fig.~1]{Modelling $\rho$ electroproduction.}
\end{figure}
%\vspace{-2mm}

As an example, let us consider a process much discussed at this meeting,
the electroproduction of vector mesons, e.g. $ep \to e \rho p$. Can we 
calculate this in strong QCD~?
\noindent The diffractive mechanism, the Pomeron part {\bf P}, is
largely phenomenological~[1], and, at
present, not amenable to calculation. Nevertheless, we can imagine modelling the process,
Fig.~1, in terms of factorisable components, some of which are
 under control~[2].
First the $\gamma^* \to q \bar{q}$~: the quarks are fully 
dressed with gluon clouds giving the whole interaction a spatial extent.
 Following this,
the quark and antiquark propagate and 
then after the Pomeron interaction form a bound state, Fig.~1.
 Importantly, each of these
 building blocks depends on the even more basic Green's function, 
 the gluon propagator,
but how are they to be calculated~?

Perhaps the best known way of performing computations in strong QCD involves the
lattice construct. However, this is not well suited to these particular problems. Firstly,
the up and down quarks, having current masses of a few MeV, do not fit on a
lattice of the size of several fermis. Moreover, as a result of
the  small $u,\,d$ quark
masses, the pion is light --- the Goldstone boson of chiral symmetry breaking.
Consequently, the emission of pions is the commonest of all hadronic processes.
However, the creation of $q \overline{q}$ pairs involves a complex
determinant on the
lattice and so this key physics is often deliberately suppressed in 
the {\it quenched}
approximation. Consequently QCD in the continuum is the natural way to study 
such problems~[3] and progress in this direction using the
Schwinger-Dyson equations is what I will briefly outline.

\begin{figure}[th]
\begin{center}
\vspace{-14mm}
\mbox{~\epsfig{file=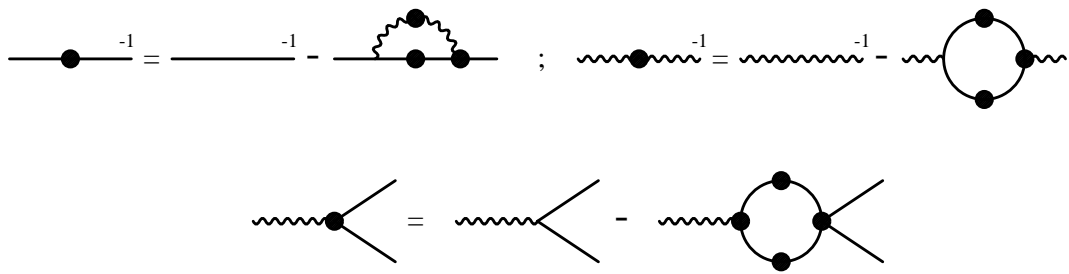,height=4.1cm}}
\end{center}
\vspace{9mm}
\caption[Fig.~2]{\leftskip 1cm\rightskip 1cm{Schwinger-Dyson equations for fermion, boson and vertex in QED.
The dots mean the Green's functions are fully dressed.}}
\end{figure}

To explain the technique and illustrate recent theoretical progress, I will 
begin by describing a slightly different strong physics problem~: dynamical
mass generation. For pedagogical simplicity I will consider QED, but we
shall shortly see this problem is closely related to chiral symmetry breaking in QCD.
We  study the field equations of the theory, the 
Schwinger-Dyson equations (SDEs). We begin with the 2-point functions, the 
fermion and boson propagators, Fig.~2. We can imagine solving these coupled
equations for the full (dressed) fermion and boson propagators --- provided we 
have an ansatz for the fermion-boson vertex. However, this 3-point function 
satisfies an SDE too relating it to 2,3 and 4-point functions and the 4-point
function satisfies its own SDE ... and so on {\it ad infinitum}.
 Consequently, the 
theory is defined by an infinite set of nested integral equations, which of
course cannot be solved without some truncation. The best understood truncation
scheme is perturbation theory, in which each Green's function is expanded in 
powers of the coupling. This not only gives a systematic procedure, but
importantly this respects gauge invariance and multiplicative renormalizability
order-by-order. However, we want to know when a fermion mass can be 
dynamically generated, even when it's bare mass is zero. This is a 
non-perturbative (strong physics) problem, since it is well-known that if the 
bare mass is zero, then the mass is zero to all orders in perturbation theory. 
We therefore need a non-perturbative truncation. It is here that progress has 
been made by recognising that ensuring gauge covariance and multiplicative
renormalisability at each level of truncation, the physics of the neglected 
higher point Green's functions is effectively pulled into the lower ones. This
is readily illustrated in quenched QED. If a crude truncation is made and the 
vertex treated as bare (the so called {\it rainbow} approximation), then it is 
well-known that a non-zero mass is generated if the interaction is strong 
enough, $\alpha \equiv e^2/4\pi > \alpha_c = \pi/3$ in the Landau gauge, as
first found by Miransky~[4]. This means a massless electron would, in the field
of a highly charged nucleus with effective coupling $Z\alpha > \pi/3$, no
longer move at the speed of light (in hypothetically quenched QED). The mass
and critical coupling should be gauge independent.
 However, explicit calculation
shows these to be strongly gauge dependent. This is a consequence of the 
unphysical truncation of the fermion SDE. If instead, one ensures that the 
vertex, the fermion-boson interaction, respects gauge covariance and
multiplicative renormalizability of the fermion propagator, a gauge independent 
mass and critical coupling occur~[5]. This marks progress in the study of the SDEs 
and ensures basic physics is retained in the truncation.
\vspace{-12mm}
\begin{figure}[hb]
\begin{center}
\mbox{~\epsfig{file=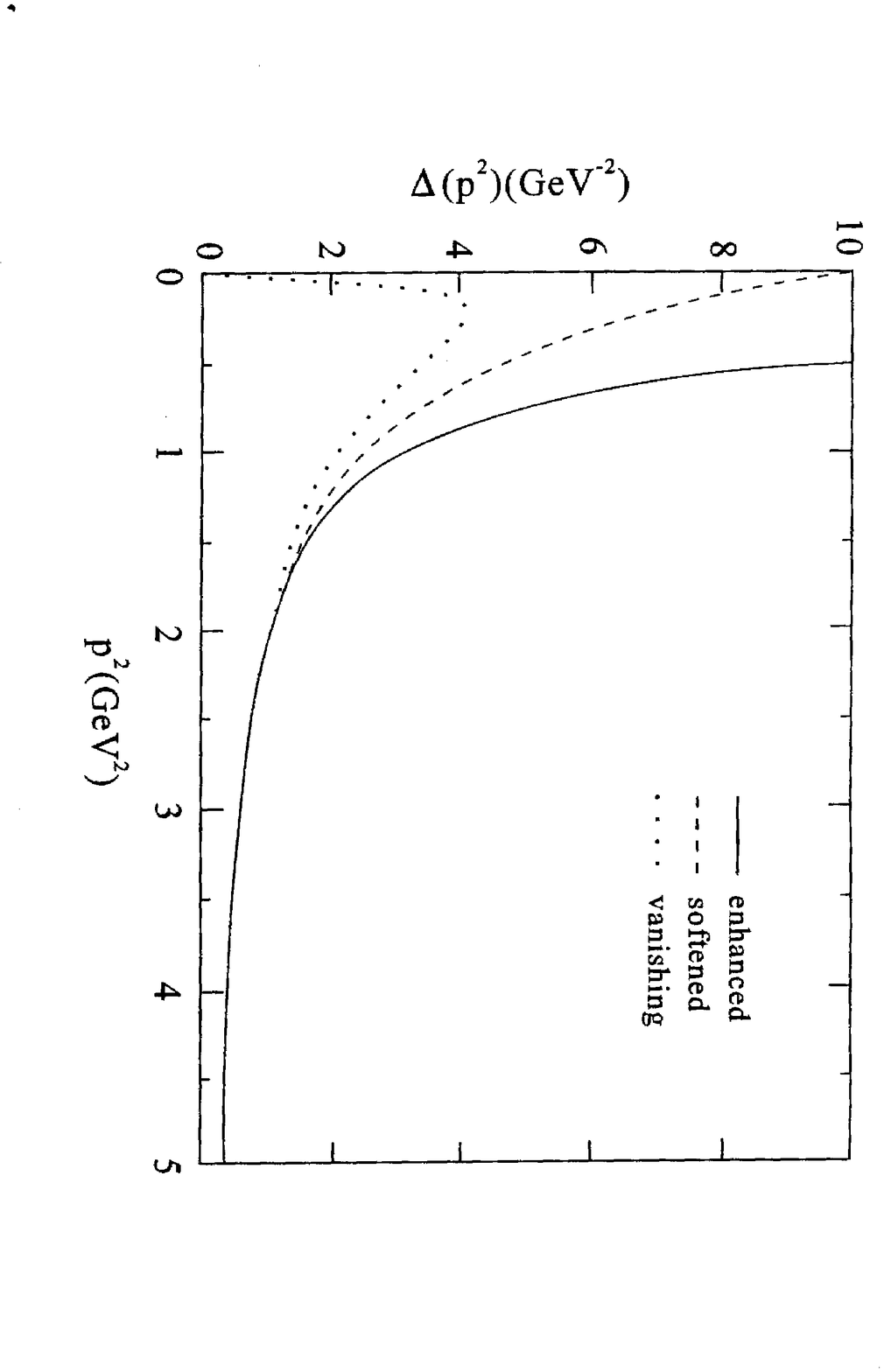,height=9.95cm,angle=90}}
%\vspace*{7cm}
\end{center}
\vspace{-7mm} 
\caption[Fig.~3]{\leftskip 1cm\rightskip 1cm{Representation of the
alternative claims for the
momentum dependence of the Landau gauge gluon propagator.}}
\end{figure}

Within this framework, we turn to QCD and its first building block, the gluon 
propagator. In covariant gauges, a full treatment is not yet possible~: the
quartic gluon coupling is suppressed and the ghosts treated perturbatively,
for example. The pure glue theory is first considered and three distinct
behaviours have been claimed for the momentum dependence of the (gauge 
variant) gluon propagator $\Delta(p^2)$. While each agrees, thanks to
asymptotic freedom, with renormalization-group improved perturbation theory
at large momenta, in the infrared they are distinct : (i) is enhanced~[6,7],
(ii) is softer than the bare propagator~[8] and (iii) is vanishing 
when $p^2 \to 0$~[9]
(Fig.~3). What has been understood in the last few years is that only the 
enhanced form is a solution of the truncated SDE. The softened behaviour is not
possible with the correct sign for the loop corrections~[10] and the infrared 
vanishing propagator occurs if the vertex has massless coloured singularities
and the Green's functions are complex in the spacelike region, where they 
should be real. Moreover, the infrared enhancement has a scale directly 
proportional to $\Lambda_{QCD}$, 
i.e. $\Delta(p^2) \sim \Lambda^2_{QCD}/p^4$~[7], so
that the stronger the interaction the smaller the size of hadrons. Such a gluon 
propagator generates a Wilson area law~[11] and does not have
 a Lehmann representation required of an asymptotic state~[3]. Consequently, the gluon
is confined by this strong self-interaction. While lattice results claim to 
support a gluon mass and not this enhanced behaviour,
they do not yet reliably probe momenta low enough to 
differentiate.

Now what effect does the infrared enhanced behaviour have on the quark 
propagator? Just as in strong QED, the interaction is sufficient to generate 
a non-zero dynamical mass~: the virtual gluon cloud gives weight to a quark,
even if it's bare mass is zero. A non-zero $\langle q\overline{q}\rangle$
condensate is created and the quark propagator is found to have no
timelike pole~: again signalling confinement~[3]. A gluon with infrared behaviour 
other than enhanced fails to produce this confinement property~[12]. While complex
singularities do arise in the momentum plane, these may be merely due to 
imperfections in the truncation or the numerical procedures and so suggest 
that the quark propagator is an entire function. This would mean that it had an
essential singularity that would prohibit the Wick rotation from Minkowski to 
Euclidean space used in many calculations here and in lattice work. This is an
issue requiring more study, but such an essential singularity may be the 
price to pay for having no free coloured states.

In principle, a complete coupling of the quark and gluon equations is really
needed and steps have been made in this direction (at least in the case of 
unbroken chiral symmetry~[13]). This would allow the determination of the $u$ and $d$ 
quark propagators in terms of $\Lambda_{QCD}$ and the current masses, $m_R$,
in some renormalization scheme. However, pending such detailed computations,
a model gluon propagator has been used with a phenomenological scale, $\kappa$,
marking the divide between the strong coupling and perturbative regimes~[14]. One
can then use the output quark propagator to build the Bethe-Salpeter (or bound
state) amplitudes for meson states by a suitable modelling of the 4-quark
kernel (something in principle determined by the SDE for the 4-point function).
To date approximating this by dressed one gluon exchange is all that has been
used. However, as shown long ago by Delbourgo and Scadron~[15], this is sufficient to
ensure that, with a non-zero $\langle q \overline{q}\rangle$--condensate, massless pseudoscalar 
bound states arise and chiral symmetry is spontaneously broken. This is unique 
to an infrared enhanced gluon. The two parameters, $\kappa$ and $m_R$, are fixed 
by the pion mass and its decay constant~[14].

We can then switch on electromagnetic interactions and calculate the pion e.m.
formfactor. Again in principle this really requires a computation of the full
$\pi \pi \gamma$ interaction. However, in practice, we may model this in the impulse
approximation by the graph of Fig.~4, in which we now know all the components,
the vertices and propagators. As shown by Craig Roberts~[16], this gives a good 
description of experiment in the spacelike region, Fig.~5, and the behaviour is 
like $\sim 1/Q^4$ (modulo logarithms) even out to $Q^2 \simeq 35$(GeV/c)$^2$. 
This beautifully highlights how non-perturbative, long range, interactions play
a role out to large momenta and explains why the perturbative result does not 
apply at present momenta.
%\newpage
\begin{figure}[th]
\begin{center}
\mbox{~\epsfig{file=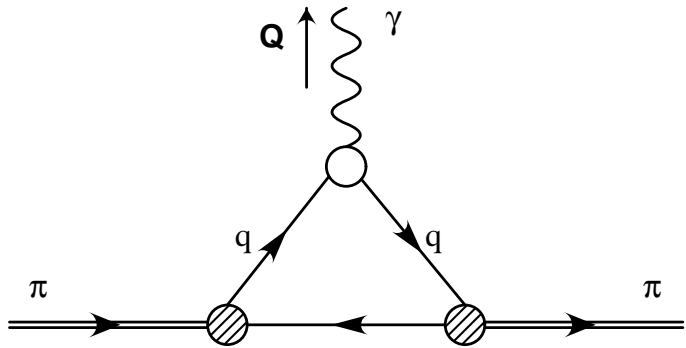,height=4.75cm}}
%\vspace*{6.5cm}
\end{center}
\vspace{-2.5mm} 
\caption[Fig.~4]{Pion electromagnetic form-factor in the impulse approximation.}
\end{figure}

We can in turn compute the $Q^2$--dependence of the pion transition form-factor 
for $\pi \gamma \to \gamma^*$, in a similar approximation. Frank et al.~[17] found
this to be in reasonable agreement with  data both old and new~[18].
In contrast, Al Mueller has emphasised~[18] that these data present an 
{\it uncontroversial} and {\it unambiguous} test of perturbative QCD.
 It is clear that data at larger
momenta, plus refinements of the calculation presented here, e.g. a more complete 
scattering kernel, as well as higher orders in perturbation theory, are all 
needed before such claims and counter-claims can be regarded as substantiated.

Lastly, we can go back to electroproduction of the $\rho$-meson. This has been
modelled by Pichowsky and Lee~[2]. The $\rho$--Bethe-Salpeter amplitude has 
parameters fixed by $\rho \to e^+ e^-$ and $\rho \to \pi \pi$ decay rates. The
agreement with the $t$ and $Q^2$--dependences is good, but clearly this is 
crucially affected by the assumed coupling of the Pomeron to dressed quarks.
The result is encouraging, but not yet a definitive prediction. The pion 
electromagnetic and transition form-factors are much more direct. They provide
an experimental probe of the infrared behaviour of the gluon that controls 
confinement and chiral symmetry breaking~: properties fundamental to the hadron 
world. This programme has a long way to go, but I hope you are convinced it
has come far.

\vskip 5mm

\noindent {\large {\bf Acknowledgements}}

\noindent I am grateful to all my colleagues in the 
{\it continuum strong physics}
community, particularly Craig Roberts, for their enthusiastic development of
this subject. Travel support was provided by the EU Training and Mobility
Programme {\it Hadronic Physics with High Energy Electromagnetic Probes}
Network FMRX-T96-0008.\\

\begin{figure}[ht]
 \begin{center}
 \begin{tabular}{ccc}
   \epsfxsize=3.95cm
   \leavevmode
   \epsffile[170 270 420 520]{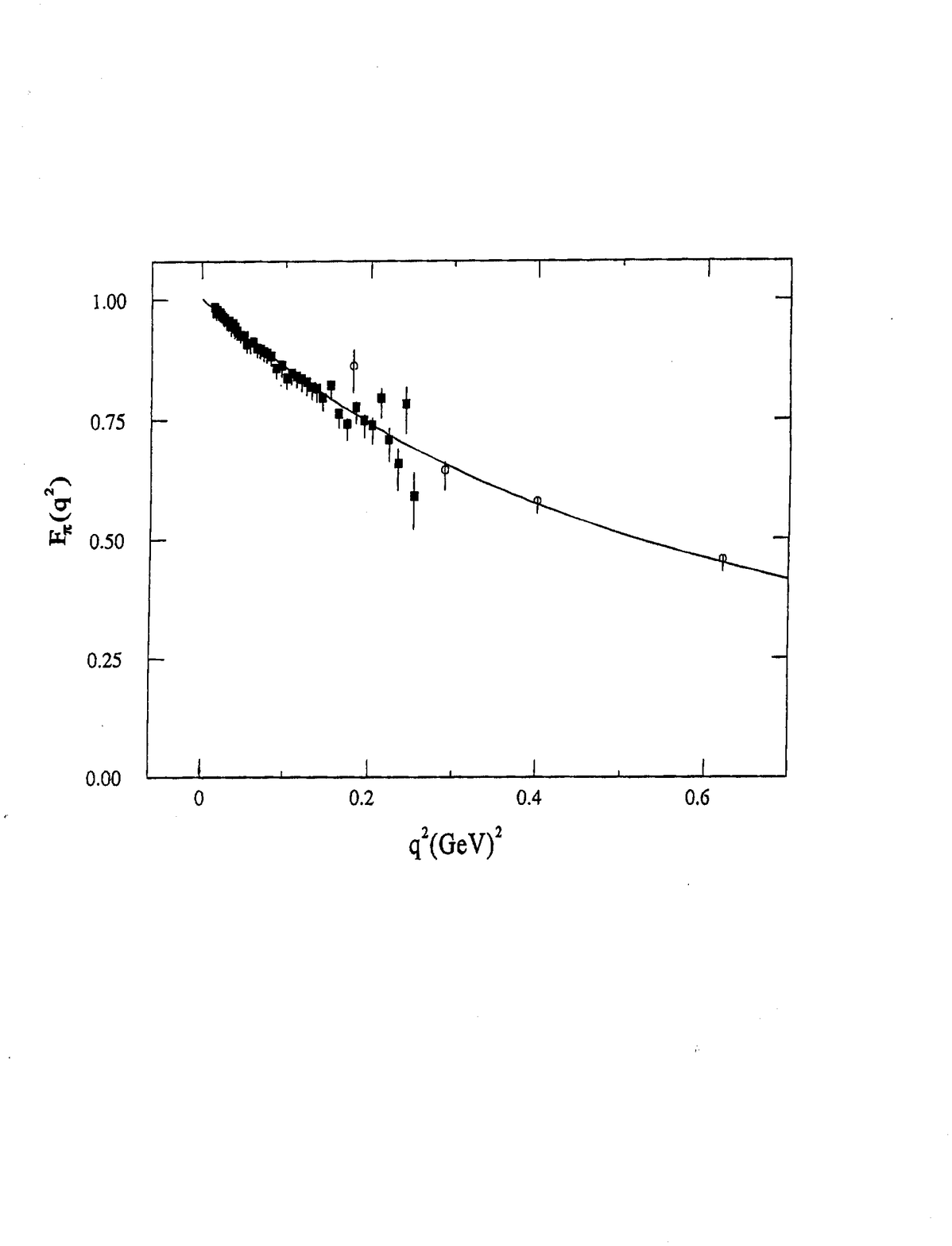}
   & \hspace{10ex} &
   \epsfxsize=3.95cm
   \leavevmode
   \epsffile[170 270 420 520]{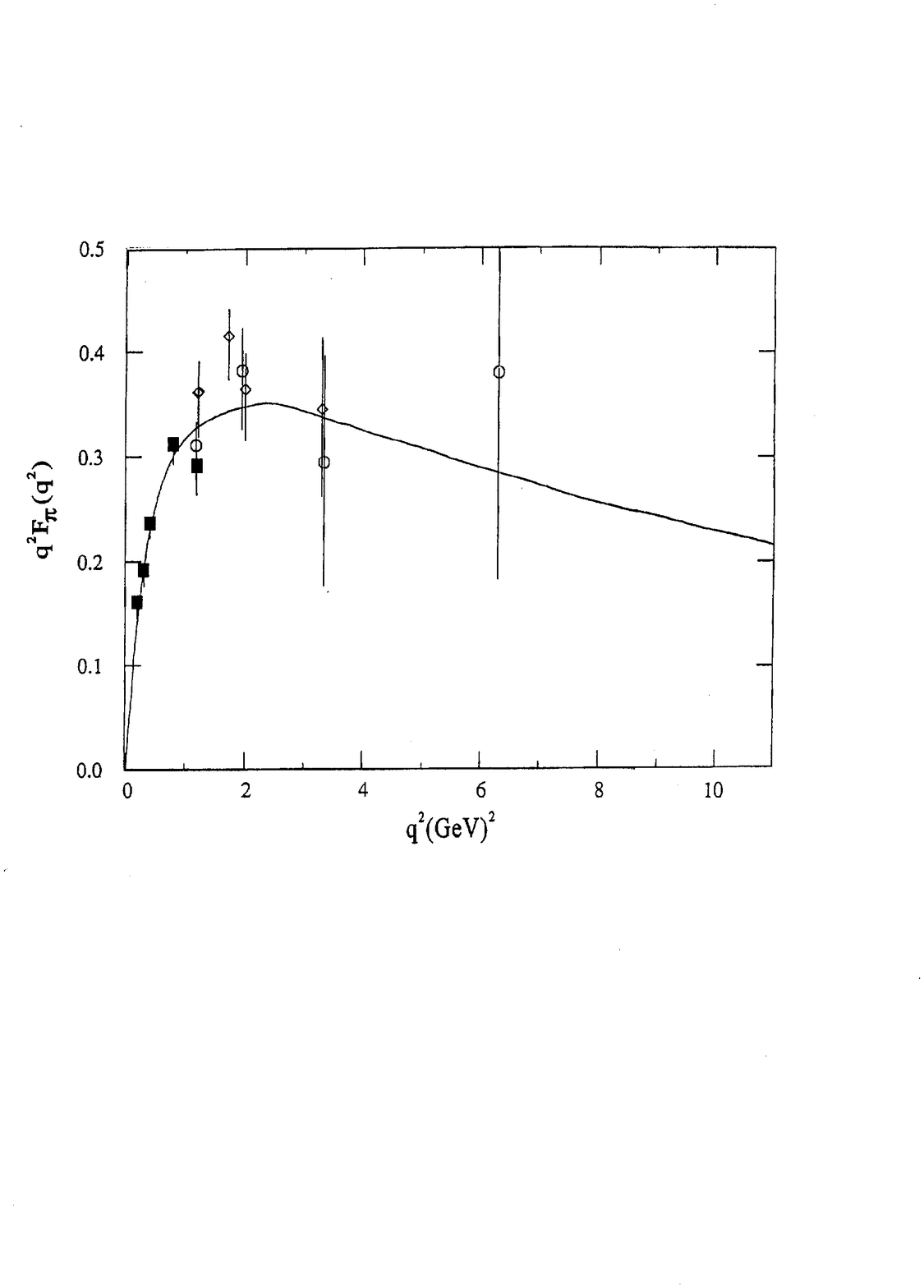}
 \end{tabular}
\vspace{-0.2cm}
\caption[Fig.~5]{Pion electromagnetic form-factor 
prediction compared with experiment [16].}
 \end{center}
\end{figure}

\noindent N.B. This list of references is far from exhaustive.
\end{document}